\documentclass[%
  prx,%
  twocolumn,%
  preprintnumbers,%
  amsmath,%
  amssymb,%
  superscriptaddress%
]{revtex4}

\usepackage{graphicx}
\usepackage{dcolumn}
\usepackage{bm}

\usepackage{xcolor}
\usepackage{soul}
\usepackage{amsmath}
\usepackage{amssymb}

\begin{document}


\title{Quantum repeater protocol for deterministic distribution of macroscopic entanglement}

\author{Alexey N. Pyrkov}
\email{pyrkov@icp.ac.ru}
\affiliation{Federal Research Center of Problems of Chemical Physics and Medicinal Chemistry RAS, Acad. Semenov av. 1, Chernogolovka, Moscow region, Russia, 142432}

\author{Ilia D. Lazarev}
\affiliation{Federal Research Center of Problems of Chemical Physics and Medicinal Chemistry RAS, Acad. Semenov av. 1, Chernogolovka, Moscow region, Russia, 142432}

\author{Tim Byrnes}
\email{tim.byrnes@nyu.edu}
\affiliation{New York University Shanghai, 567 West Yangsi Road, Shanghai, 200126, China; NYU-ECNU Institute of Physics at NYU Shanghai, 3663 Zhongshan Road North, Shanghai 200062, China; Shanghai Frontiers Science Center of Artificial Intelligence and Deep Learning, NYU Shanghai, 567 West Yangsi Road, Shanghai, 200126, China.}
\affiliation{State Key Laboratory of Precision Spectroscopy, School of Physical and Material Sciences, East China Normal University, Shanghai 200062, China}
 \affiliation{Center for Quantum and Topological Systems (CQTS), NYUAD Research Institute, New York University Abu Dhabi, UAE.}
\affiliation{Department of Physics, New York University, New York, NY 10003, USA}

\date{\today}

\begin{abstract}
Distributing long-distance entanglement is a fundamental goal that is necessary for a variety of tasks such as quantum communication, distributed quantum computing, and quantum metrology.  Currently quantum repeater schemes typically  aim to distribute one ebit at a time, the equivalent of one Bell pair's worth of entanglement.  Here we present a method to distribute a macroscopic amount of entanglement across long-distances using a number of operations that scales only linearly with the chain length.  The scheme involves ensembles of qubits and entangling them with an $S^z S^z$ interaction, which can be realized using atomic gas ensembles coupled by a shared optical mode. Using only local measurements on the intermediate ensembles, this leaves the ensembles at the ends of the chain entangled.  We show that there are particular ``magic'' interaction times that allow for distribution of entanglement with perfect fidelity, with no degradation with chain length.  The scheme is deterministic, such that with suitable local conditional unitary corrections, the same entangled state can always be prepared with good approximation.  
\end{abstract}

\maketitle

\section{Introduction}

Entanglement distribution in quantum networks is an important task in the field of quantum information, with potential applications such as secure communication, quantum metrology, distributed quantum computation \cite{Wei2022,Wehner2018,Azuma2023}. Photons can carry quantum information over long distances with almost negligible decoherence and are compatible with existing telecommunication fiber technology.  
However, while photons can be transferred relatively efficiently between two locations in the same laboratory or city, distances exceeding this begin to be problematic due to photon loss. 
For example, in a typical optic fiber, the photon loss rate become significant beyond distances of $ \sim 200 $ km and the maximum distance is estimated as $\sim 600$ km \cite{Pirandola2017, Yingqiu2021}. 
In classical signal transmission line this problem can be solved with optical amplifiers but it is not possible to amplify quantum signal due to the famous no-cloning theorem \cite{Park1970, Wootters1982, Nielsen2010}. To overcome this difficulty, Briegel, Zoller and colleagues proposed the quantum repeater protocol \cite{Briegel1998}, 
aiming to establish long-distance quantum entanglement. 
In the quantum repeater scheme, the transmission channel is divided into several short-distance links, where entanglement is generated between the nearest neighbor nodes using photon transmission.  Then entanglement between distant nodes is achieved through entanglement swapping \cite{bennett1993teleporting,zukowski1993event,Pan1998}. 

In the entanglement swapping protocol, the probabilistic nature of generating photonic entanglement (e.g. by parametric down conversion) across the elementary links means that the entanglement may not be present synchronously.  One potential solution to this challenge involves storing the generated entanglement in a quantum memory until it has been successfully prepared in both links \cite{Lvovsky2009, Simon2010, Heshami2016}. It has been shown that many physical systems, including single neutral atoms \cite{Reiserer2015,Rosenfeld2008}, 
cold atoms \cite{Sangouard2011}, 
trapped ions \cite{Duan2010}, 
and the Nitrogen-vacancy centers in diamonds \cite{Childress2006, Gurudev2007,hermans2022qubit}  
can absorb and store incoming photons, 
for instance using techniques such as electromagnetically induced transparency \cite{Fleischhauer2005}.  This allows them to serve as the nodes of a quantum network with an embedded quantum memory \cite{Duan2001,Duan2010,Ritter2012}. 

As a platform to realize a quantum memory, atomic gas ensembles are an attractive candidate due to their ability to perform strong and controllable coupling between the atoms and photons.  The primary advantanges of using an ensemble is that there is a collective enhancement due to the large number of atoms with long coherence times. Duan, Lukin, Cirac, and Zoller (DLCZ) proposed a scheme based on single-photon detection to implement a quantum repeater and realize scalable long-distance quantum communication with atomic ensembles \cite{Duan2001}. 
In the scheme, probabilistic write-out photons are generated from two remote atomic ensembles, and the correlated photons are sent to a beam splitter generating a heralded Fock-state type entanglement between the atomic ensembles. Kuzmich, van der Wal, and co-workers experimentally realized a quantum memory using the DLCZ protocol with trapped cold Cs atoms \cite{Kuzmich2003} and Rb atoms \cite{Wal2003}. 
This was followed by several other works studying entanglement generation between atomic ensembles  \cite{Chou2005,Yuan2008,Yu2020,Pu2021,Simon2007,Bao2012,Sangouard2008,liu2024creation}.  In the continuous variable (CV) framework, Polzik produced entanglement between atomic ensembles using a quantum nondemolition measurement approach \cite{julsgaard2001experimental,krauter2013deterministic}. Another prominent design for a CV repeater involves creating entangled two-mode quadrature squeezed states of light using spontaneous parametric downconversion (SPDC) sources. This is followed by CV non-Gaussian quantum error correction through the injection and detection of single photons \cite{Dias2018}. The process also includes storing heralded link-level entanglement in quantum memories, as well as entanglement swapping through coherent detection \cite{Dias2020,Ghalaii2020,Furrer2018} or other non-Gaussian methods \cite{Seshadreesan2020}.


Such approaches for generating entanglement between ensembles work in either the single excitation  (e.g.  DLCZ) or Holstein-Primakoff regime (for CV), such that the amount of entanglement is of the order of one ebit, corresponding to the entanglement in a single Bell pair.  However, the very large number of atoms in an atomic ensemble provides an opportunity to create entanglement in a different regime, creating a type of {\it macroscopic} entanglement.  Such an approach was discussed in works such as Refs. \cite{byrnes2013fractality,pyrkov2013entanglement,rosseau2014,byrnes2012macroscopic,hussain2014geometric}, where a scheme for producing macroscopic entanglement using an $ S^z S^z$ interaction was analyzed and proposed to be used for quantum computation. Here, we define macroscopic entanglement as von Neumann entropy or logarithmic negativity that is of order of $ \log_2 D $, where $ D \gg 1  $ is the dimensionality of the subsystems that participate in the entanglement.  In the case of atomic ensembles, the effective dimensionality is equal to the number of atoms \cite{byrnes2020quantum}, which is in the range of $ 10^3 $ to $ 10^{12} $.   These type of states have been proposed to be used as a resource for several quantum information processing protocols, such as quantum teleportation, remote state preparation, and others \cite{pyrkov2014full,byrnes2015macroscopic,manish2021,byrnes2011accelerated}. Experimentally, 
the creation of many-particle entanglement localized in a single spatial location \cite{schmied2016bell} and spatially separate regions \cite{fadel2018spatial} within one Bose-Einstein condensate (BEC) has been observed. 
Bell correlations have been experimentally confirmed in a BEC \cite{schmied2016bell} and a thermal atomic ensemble \cite{engelsen2017}. Recently, an experiment successfully demonstrated the achievement of entanglement between a spatially split BECs \cite{Colciaghi2023}. 

\begin{figure}[t]
\centering
\includegraphics[width=\columnwidth]{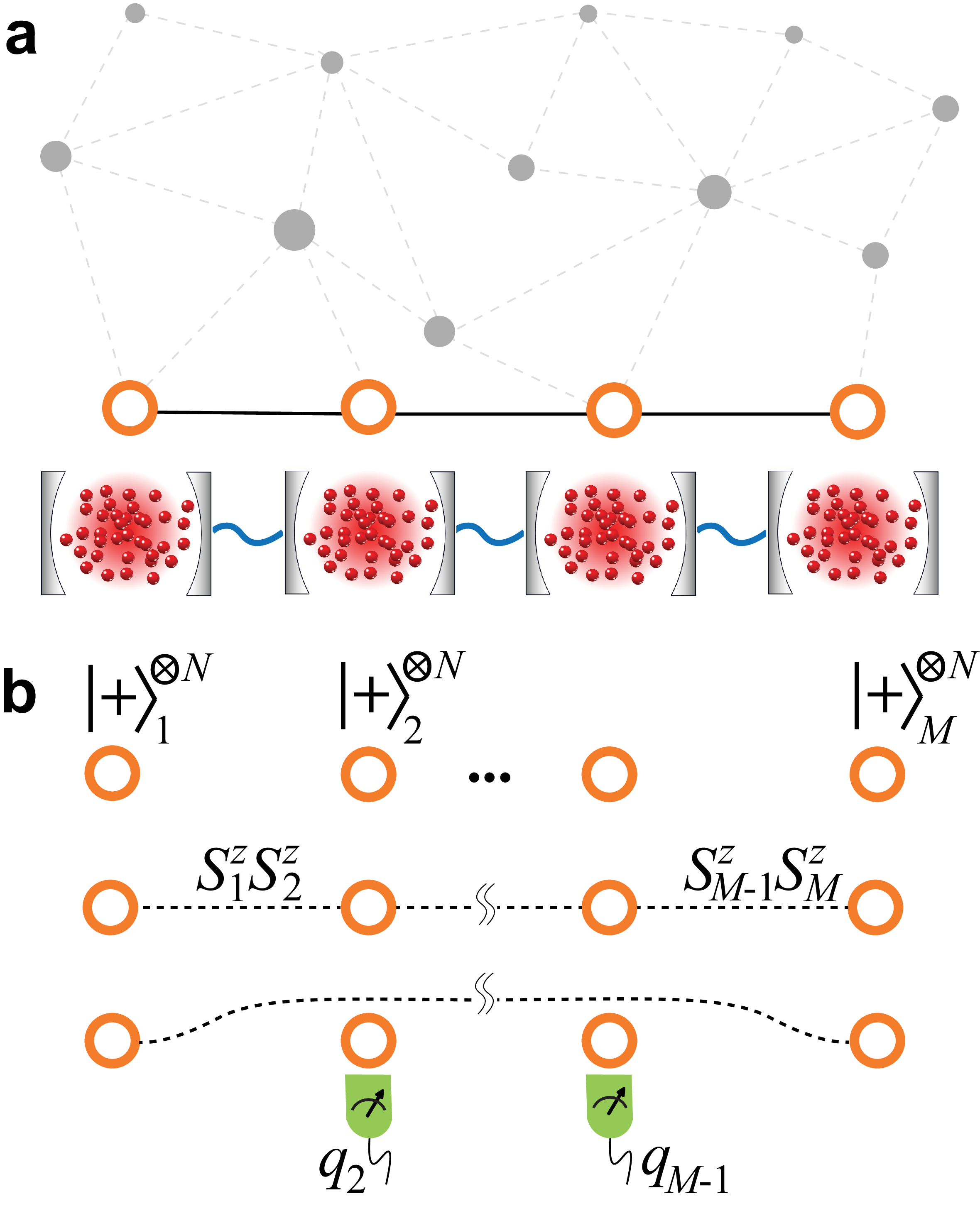}
\caption{\label{fig1}
Macroscopic entanglement distribution with qubit ensembles.  (a) Ensembles of qubits are placed at the nodes of a quantum network.  The nearest neighbors of the network are capable of being entangled with a $ S^z S^z $ Hamiltonian, which can generate macroscopic entanglement.  The aim is to create entanglement between distant nodes of the network.  A possible physical implementation is using atomic ensembles placed in cavities connected by an optical fiber, such as discussed in Ref. \cite{pyrkov2013entanglement}. 
(b) The entanglement distribution protocol.  First the ensembles at each node are polarized in the $ S^x $ direction, thus preparing in the initial state (\ref{initstate}).  Here we write $ | + \rangle^{\otimes N } = | \frac{1}{\sqrt{2}}, \frac{1}{\sqrt{2}} 
\rangle \rangle $. Nearest neighbor nodes are entangled with the Hamiltonian (\ref{mainham}), producing the state (\ref{evenpsiM}).  Finally the intermediate nodes are measured, giving the state (\ref{final_even}). Dashed lines connecting the nodes denote the entanglement.    }
\end{figure}

In this paper, we introduce a new protocol for creating long-distance {\it macroscopic} entanglement between atomic ensembles. The basic idea is to use entanglement swapping, in a similar way to that proposed in existing quantum repeaters, but perform this at the macroscopic scale. As illustrated in Fig. \ref{fig1}, we consider a chain of qubit ensembles within a quantum network, where nearest neighbors can be entangled with a $S^z S^z$ interaction.  Only local projective measurements are required on the intermediate ensembles leaving the ensembles at the ends of the chain in an entangled state (Fig. \ref{fig1}(b)).  Operations at the ensemble level are only performed for the protocol (i.e. no microscopic operations of single qubits), and the resources required for the protocol only depends upon the chain length and not the number of qubits in an ensemble. 
The type of state is considerably more complex than the qubit version due to the many-body entangled state created by the $S^z S^z$ interaction between the atomic ensembles. Despite this, we are able to find an analytic solution at particular entanglement times.  We show that there are ``magic'' interaction times  where entanglement distributed to the ends of the chain do not degrade with the length of the chain.

\section{Entanglement distribution protocol}

We now describe concretely the entanglement distribution protocol using qubit ensembles. The basic overview of the scheme is shown in Fig. \ref{fig1}(b).  In order to keep our presentation independent of a particular physical implementation, we consider generically a system consisting of $ N $ qubits located at each node of the quantum network.  We assume that the qubits within the ensemble cannot be controlled individually, but the ensembles may be controlled by total spin operations as we describe below. 

Each of the ensembles is initially prepared in a spin coherent state, which we denote 
\begin{equation}
\label{ensemblequbit}
|\alpha,\beta\rangle\rangle\equiv \prod_{l=1}^N (\alpha |0\rangle_l +\beta |1 \rangle_l).  
\end{equation}
where $ |0\rangle_l $ and $ |1\rangle_l $ denotes the computational states for the $l$th qubit in the ensemble.  Here, $\alpha$ and $\beta$ are arbitrary complex numbers satisfying $|\alpha|^2+|\beta|^2=1$. 

Each ensemble is controlled using total spin operators
\begin{align}
S^\gamma = \sum_{l=1}^N \sigma_l^\gamma
\end{align}
where $\gamma \in\{ x, y, z \} $ and $ \sigma_l^\gamma $ is the Pauli spin operator for the $ l $th qubit within the ensemble.  The spin operations conserve the particle number, and we assume that $ N $ is the same for each ensemble for simplicity. 

Since the spin operators and the initial state are completely symmetric under particle interchange, one can write an equivalent representation using the Schwinger boson formalism \cite{byrnes2020quantum}.  The spin coherent state in the bosonic formulation is 
\begin{equation}
\label{becqubit}
|\alpha,\beta\rangle\rangle\equiv\frac{1}{\sqrt{N!}}(\alpha a^\dagger+\beta b^\dagger)^{N}|0\rangle,
\end{equation}
where the bosonic creation and annihilation operators $ a, b $ obey commutation relations $[a,a^\dagger]=[b,b^\dagger]=1$. The spin operators are written as
\begin{align}
S^x & = a^\dagger b + b^\dagger a \nonumber \\
S^y & = -i a^\dagger b + i b^\dagger a \nonumber \\
S^z & = n^a - n^b \nonumber \\
\hat{N} & = n^a + n^b , 
\end{align}
where the number operators for the two qubit states are defined as $ n^a = a^\dagger a, n^b = b^\dagger b $.  The eigenstates of $ S^z $ operator, which we call the number states, are defined as
\begin{align}
| k \rangle = \frac{ (a^\dagger)^k (b^\dagger)^{N-k} }{\sqrt{k! (N-k)!}} | \emptyset \rangle ,
\end{align}
where $ | \emptyset \rangle $ is the vacuum state.  The eigenvalues are $ S^z |k \rangle = (2k - N) | k \rangle $. A mapping of the number states in the qubit formulation is given in Ref. \cite{byrnes2023multipartite}.



We consider a chain of $ M $ ensembles prepared in the initial state
%
%
\begin{equation}
|\phi_0 \rangle=|\frac{1}{\sqrt{2}},\frac{1}{\sqrt{2}}\rangle\rangle_{1}\otimes\dots\otimes|\frac{1}{\sqrt{2}},\frac{1}{\sqrt{2}}\rangle\rangle_{M} .  \label{initstate}
\end{equation}
We assume a scenario where the nearest neighbors of the chain can be entangled using a $ S^z S^z $  interaction. More distant pairs cannot be directly entangled.  The Hamiltonian that we will consider in the chain is
\begin{align}
H=  \sum_{j=1}^{M-1} (-1)^j n_j^a n_{j+1}^a , 
\label{mainham}
\end{align}
where $ j$ labels the  $M $ ensembles. The signs in the Hamiltonian are chosen in this way so that the simplest analytical formulas can be obtained, as shown later.  
 This is the same as a $ S^z S^z $ interaction up to local rotations because
\begin{align}
n^a_j n_{j+1}^a = \frac{1}{4} (S^z_j + \hat{N}_j) (S^z_{j+1} + \hat{N}_{j+1} ).  
\end{align}

Numerous proposals for producing a $ S^z S^z $ interaction between qubits have been proposed, particular in the context of atomic ensembles and BECs.  In Ref. \cite{pyrkov2013entanglement} a scheme mediated by photons in an optical fiber was proposed.  This was later adapted to a geometric phase gate \cite{hussain2014geometric}. Another approach is to use spin-dependent forces on neighboring BECs \cite{treutlein2006}.  
Such an interaction is also produced when one-axis squeezed states are split \cite{Jing_2019}.  This type of entanglement was  experimentally realized in BECs \cite{fadel2018spatial,Colciaghi2023}.  In this paper, we assume that such an interaction can be induced using suitable methods, and the interaction time $ t $ is a controllable parameter. 

Evolving the Hamiltonian for a time $ t $ between the ensembles gives the unitary evolution
\begin{equation}
U=\prod^{M-1}_{j=1} e^{- i (-1)^j  n^a_j n^a_{j+1} t} . 
\end{equation}
The state after applying the gates is
\begin{align}
|\psi_M \rangle = U | \phi_0 \rangle . 
\end{align}

To understand the type of state that is produced by this interaction, consider first the $ M = 2$ case which gives
\begin{align}
|\psi_2 \rangle&=e^{i n^a_1 n^a_{2} t}|\frac{1}{\sqrt{2}},\frac{1}{\sqrt{2}}\rangle\rangle_{1}|\frac{1}{\sqrt{2}},\frac{1}{\sqrt{2}}\rangle\rangle_{2}\nonumber\\
&=\frac{1}{\sqrt{2^N}}\sum_{k_1}\sqrt{C_{N}^{k_1}}|k_1\rangle |\frac{e^{i k_1 t} }{\sqrt{2}},\frac{1}{\sqrt{2}}\rangle\rangle_2 . 
\label{M2psi}
\end{align}
As discussed in Ref. \cite{byrnes2013fractality}, this produces a type of entangled state where number states in the first ensemble are correlated with spin coherent states around the equator of the Bloch sphere of the second.  The unitary operation $U $ can be viewed as a conditional rotation of ensemble 2 with an angle dependent on the number state on ensemble 1.   

%




For a chain of $ M $ ensembles, the wavefunction after the entangling gate is 
\begin{multline}
|\psi_M \rangle=\frac{1}{\sqrt{2^{\frac{MN}{2}}}}\sum_{k_1,k_3,\ldots,k_{M-1}=0}^{N} \left( \prod_{i \in \text{odd}} \sqrt{ C^{k_i}_{N}}  \right) \\
\times | k_1 \rangle_1  |\frac{e^{i(k_{1}-k_{3})t}}{\sqrt{2}},\frac{1}{\sqrt{2}}\rangle\rangle_2 | k_3 \rangle_3
|\frac{e^{i(k_{3}-k_{5})t}}{\sqrt{2}},\frac{1}{\sqrt{2}}\rangle\rangle_4  \\
\dots \otimes  | k_{M-1} \rangle_{M-1} |\frac{e^{ik_{M-1}t}}{\sqrt{2}},\frac{1}{\sqrt{2}}\rangle\rangle_{M}  
\label{evenpsiM}
\end{multline}

%
%
for even $ M $. We have 
\begin{multline}
| \psi_M \rangle=\frac{1}{\sqrt{2^{\frac{(M+1)N}{2}}}}\sum_{k_1,k_3,\ldots,k_M=0}^{N}\left( \prod_{i \in \text{odd}} \sqrt{ C^{k_i}_{N}}  \right) \\
\times | k_1 \rangle_1  |\frac{e^{i(k_{1} - k_{3})t}}{\sqrt{2}},\frac{1}{\sqrt{2}}\rangle\rangle_2 | k_3 \rangle_3
|\frac{e^{i(k_{3} - k_{5})t}}{\sqrt{2}},\frac{1}{\sqrt{2}}\rangle\rangle_4  \\
\dots \otimes |\frac{e^{i(k_{M-2} - k_{M})t}}{\sqrt{2}},\frac{1}{\sqrt{2}}\rangle\rangle_{M-1} | k_{M} \rangle_{M} 
\label{oddpsiM}
\end{multline}
%
%
%
for $M$ is odd.  In the expressions (\ref{evenpsiM}) and (\ref{oddpsiM}), the odd numbered ensembles are expanded to number states and the remaining ensembles on even sites are conditionally rotated. We note that the wavefunction can be always written in a way such that on adjacent sites, there is a conditional rotation of the ensemble written as a spin coherent state, dependent on the number states on alternating nodes.  Hence the parity of the number of ensembles $ M $ plays important role.  Depending on the parity of $ M $, the last ensemble is written in terms of a spin coherent state or a number state.  Henceforth, we will show expressions for even $ M $ and defer the expressions for odd $ M $ to Appendix \ref{sec:oddMexpressions}.


Next we perform a measurement of the intermediate ensembles in the $x$-basis.  The measurement operator for this is written 
\begin{align}
M_q^{(j)} = | q \rangle^{(x)}_j \langle q |^{(x)}_j ,
\label{measurementop}
\end{align}
where the number state in the $ x $-basis is $ | k  \rangle^{(x)} = e^{-iS^y \pi/ 4} | k  \rangle $ and $ q \in [0,N] $ labels the measurement outcome. 
We note that in conventional entanglement swapping, two qubits from different Bell pairs are measured in the Bell basis to create long-distance entanglement. Such entangled basis measurements are implementable in a photonic context but are less convenient in the context of atomic ensembles.  Here, we perform only local measurements on intermediate nodes such as to transfer the entanglement between the first and last ensembles in the chain.  

Later, we show that it is beneficial to introduce a slight angular offset from the $ x $-axis of the measurement.  For this purpose we perform a equatorial rotation before performing the measurement, given by 
\begin{align}
V^{(j)}(\phi)  = e^{i n^a_j \phi} .
\label{equatorrot}
\end{align}
Hence the final state is the projected state
\begin{align}
| \Psi_{\vec{q}} \rangle = \left( \bigotimes_{j=2}^{M-1} M^{(j)}_{q_j} V^{(j)}(\phi) \right) | \psi_M \rangle  ,
\end{align}
which is an unnormalized state.  Here we wrote the full set of measurement outcomes of the intermediate ensembles as
\begin{align}
\vec{q} = (q_2, q_3, \dots, q_{M-1} ) .
\end{align}

The final unnormalized state after all the measurements is 
\begin{multline}
| \Psi_{\vec{q}} \rangle =\frac{1}{\sqrt{2^{\frac{MN}{2}}}}\sum_{ k_1, k_3, \dots, k_{M-1}  = 0}^{N}  \sqrt{ C^{k_1}_{N}} \left( \prod_{j=2}^{M-1} \Omega_{q_j}^{(j)} \right)
  \\ \times  |k_1\rangle|\frac{e^{i k_{M-1}t}}{\sqrt{2}},\frac{1}{\sqrt{2}}\rangle\rangle_M 
\label{final_even}
\end{multline}
%

%
where we defined
\begin{align}
\Omega_{q}^{(j)} = \left\{
\begin{array}{cc}
\langle q |^{(x)} | \frac{e^{i(k_{j-1} - k_{j+1})t + i \phi }}{\sqrt{2}},\frac{1}{\sqrt{2}}\rangle\rangle
 & j \in \text{even} \\
e^{i k_j \phi} \sqrt{ C^{k_j}_{N} } \langle q|^{(x)} | k_j \rangle &  j \in \text{odd}
\end{array}
\right.   . 
\end{align}
The explicit expressions for the matrix elements are 
\begin{equation}
\langle q |^{(x)} | \frac{e^{i\alpha}}{\sqrt{2}},\frac{1}{\sqrt{2}}\rangle\rangle = i^{N-q} e^{i N  \alpha /2} \sqrt{C_N^{q}}\cos^{q} \frac{\alpha}{2} \sin^{N-q} \frac{\alpha}{2} 
\end{equation}
and
\begin{multline}
\langle q|^{(x)} | k\rangle=\frac{1}{\sqrt{q!(N-q)!2^N}}\sum^q_{l=0}\sum^{N-q}_{m=0}C^l_q C^m_{N-q}\\
(-1)^{N-q-m}\sqrt{(l+m)!(N-l-m)!}\delta_{k,l+m}
\end{multline}
where $\delta_{i,j}$ is Kronecker delta \cite{byrnes2020quantum}. The probability of obtaining this outcome labeled by $ \vec{q} $ is
\begin{align}
p_{\vec{q}} = \langle \Psi_{\vec{q}} | \Psi_{\vec{q}} \rangle . 
\end{align}

\begin{figure}
    \centering
    \includegraphics[width=\columnwidth]{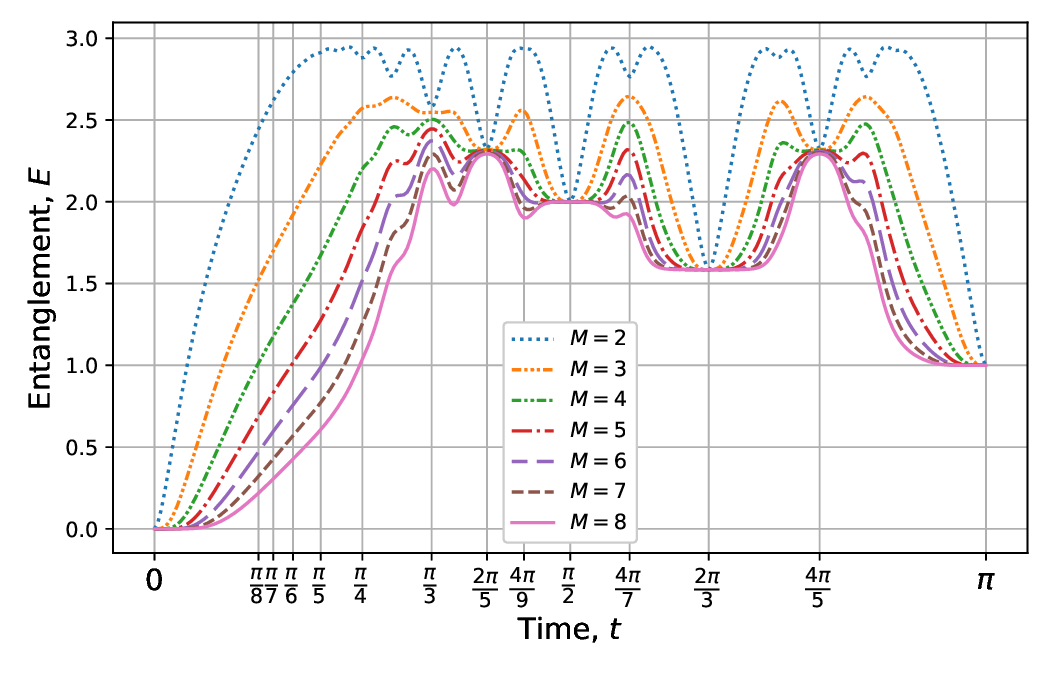}
    \caption{Entanglement as measured by von Neumann entropy (\ref{entropy})  between the first and last nodes of the chain, after projecting out the intermediate ensembles. The states used are (\ref{final_even}) for even $ M $ and (\ref{final_odd}) for odd $ M $.  The measurement outcome $ q_j = N $ is used for all curves with offset angle $ \phi = 0 $.  The interaction time $ t $ is with respect to the Hamiltonian (\ref{mainham}) and different lines show various chain lengths $ M $. Each ensemble has $ N = 20 $ bosons. }
    \label{fig:entanglement}
\end{figure}

\section{Entanglement distribution}

The entanglement for the states can be calculated using the von Neumann entropy
\begin{equation}
E=-\sum^{N}_{l=0} \lambda_l \log_2 \lambda_l ,
\label{entropy}
\end{equation}
where $\lambda_l$ are the eigenvalues of the density matrix
\begin{equation}
\rho_M = \frac{\text{Tr}_1  |\Psi_{\vec{q}} \rangle\langle\Psi_{\vec{q}} |}{p_{\vec{q}}} =
\frac{1}{p_{\vec{q}}} \sum^N_{k_1=0} \langle k_1|\Psi_{\vec{q}} \rangle\langle \Psi_{\vec{q}} |k_1\rangle . 
\label{rhom}
\end{equation}

The dynamics of entanglement versus time and the number of nodes in chain is shown in Fig. \ref{fig:entanglement}. Here we show the case when all measurement results are chosen equal to $q_j = N$ and the offset angle $ \phi = 0 $. We later generalize our analysis to other measurement outcomes, but this case shall be an interesting particular case to examine first.  We firstly note that the form of the graph is similar to the characteristic ``devil's crevasse'' shape giving sharp dips at fractional multiples of $ \pi $ \cite{byrnes2013fractality}.  The fact that this type of dependence is seen can be understood from the fact that the underlying entanglement between nearest neighbor nodes is generated using  $ S^z S^z $ interactions.  It naturally follows that the more entanglement that is present between the neighboring nodes, the larger the amount of entanglement that is distributed to the ends of the chain after the projection is applied.  

An important question is the performance with scaling the protocol with chain length $ M $, such that long distance entanglement distribution can be achieved.  Typically, we observe from Fig. \ref{fig:entanglement} that the entanglement degrades with increasing chain length (see for example short interaction times $t < \pi/4 $).  What is interesting is that for some particular times the entanglement does not degrade, and remains constant regardless of the chain length. We call these special points in the graph ``magic times'', as these regions are of particular interest in the context of entanglement distribution.  It is also interesting to see that as the chain length increases, in the vicinity of the magic times the entanglement curve flattens out and remains constant for a range of times beyond the magic times.  This is important as it suggests that there is some tolerance allowable for the interaction times, which is beneficial if some errors exist in controlling $ t $.

\begin{figure}
    \centering
    \includegraphics[width=\columnwidth]{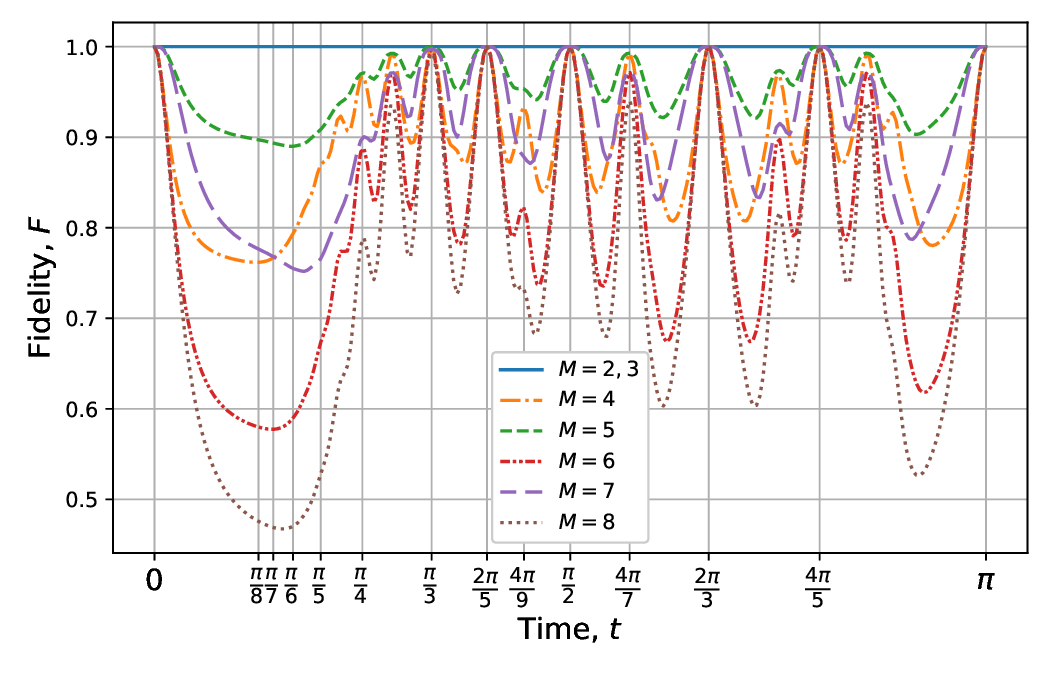}
    \caption{The fidelity between the projected state $ | \Psi_{\vec{q}} \rangle  $ and the $ M = 2$ state (\ref{M2psi}) for even cases and the $ M = 3$ state (\ref{ideal_odd}) for odd cases.  Expressions for the fidelity are given by 
       (\ref{fideven}) for even $ M $ and (\ref{fidelity_odd}) for odd $ M $.  The measurement outcome $ q_j = N $ is used for all curves with offset angle $ \phi = 0 $.  The interaction time $ t $ is with respect to the Hamiltonian (\ref{mainham}) and different lines show various chain lengths $ M $. Each ensemble has $ N = 20 $ bosons. }
    \label{fig:fidelity}
\end{figure}

To understand the nature of the state that is generated at the magic times,  we evaluate the fidelity with respect to the $ M = 2$ state wavefunction given by (\ref{M2psi}). The fidelity is given by 
\begin{multline}
F = \frac{| \langle  \psi_2 | \Psi_{\vec{q}} \rangle |^2}{p_{\vec{q}}}  = \frac{1}{2^{N(M+2)/2} p_{\vec{q}} } \\
\times \Big| \sum_{k_1, k_3, \dots, k_{M-1} }C_N^{k_1}
\left( \prod_{j=2}^{M-1} \Omega_{q_j}^{(j)} \right) \left(\frac{e^{i(k_{M-1} - k_1)t }+1}{2}\right)^N \Big|^2 .  
\label{fideven}
\end{multline}
For an odd number of nodes in Fig. \ref{fig:fidelity}, we evaluate the fidelity with respect to the $ M = 3$ state wavefunction
\begin{equation}
    \label{ideal_odd}
| \Psi_N^{(M=3)}  \rangle=\frac{\sum_{k_1,k_3=0}^{N} \sqrt{ C^{k_1}_{N}  C^{k_3}_{N} } 
 \Omega_{N}^{(2)}  |k_1\rangle|k_3\rangle }{\sqrt{\sum_{k_1,k_3=0}^{N}  C^{k_1}_{N}  C^{k_3}_{N} | \Omega_{N}^{(2)} |^2 } } .
\end{equation}
Explicit expressions for the fidelity are given in Appendix \ref{sec:oddMexpressions}.  The fidelity versus the interaction time is shown in Fig. \ref{fig:fidelity} for various chain lengths. We see that at the magic times the fidelity of the state is equal to 1, for all chain lengths.  This shows that the two types of states corresponding to the $ M = 2 $ state Eq. (\ref{M2psi}) the $ M = 3 $ state Eq. (\ref{ideal_odd}) can be prepared at these magic times without degradation with the length of the chain. 

These results suggest that it is possible to distribute the same type of macroscopic entanglement as that generated using the $ S^z S^z $ interaction, as long as the interaction times are in the vicinity of the magic times.  However, for this to be considered a practical protocol, we must also analyze outcomes beyond $ q_j = N $, as this is only a single measurement outcome from many. 
For long chain lengths, there is an exponentially small probability of obtaining $ q_j = N $, and  would be extremely inefficient if only this was the only successful outcome.  If one simply considers other measurement outcomes we do not have the stationary entanglement property as seen in Fig. \ref{fig:entanglement}, and the particular state that is generated is dependent on the particular measurement outcome.  We now show that by modifying the measurement slightly it is possible to produce the same entanglement between the end nodes of the chain for any measurement outcome.

\section{Spin-cat states}

To understand what is special about the entanglement at the magic times we perform an analysis of the wavefunction at these times.  For the purposes of our analysis we define the magic times to be at evolution times 
\begin{align}
t_L = \frac{2\pi}{L}
\label{magictimes}
\end{align}
where $L $ takes an integer value. We note that there are additional times that have the invariance property with $ M$, e.g. $ t = 4\pi/5 $ (see Fig. \ref{fig:fidelity}). For the simplicity and generality of the analysis we shall only consider the times (\ref{magictimes}).  For all times  $ L \in [1, \sqrt{N} ] $ the entanglement exhibits the stationary behavior.  

The reason why times with fractional multiplies of $ 2 \pi  $ are special is discussed in Ref.  \cite{byrnes2013fractality} for the $M = 2 $ case.  At these times we see that the the spin coherent states factorize as
\begin{align}
&|\psi_2 (t= \frac{2\pi}{L} )  \rangle=\frac{1}{\sqrt{2^N}}\sum_{k=0}^N \sqrt{C_{N}^{k}}|k\rangle_1  |\frac{e^{2 \pi i k/L }}{\sqrt{2}},\frac{1}{\sqrt{2}}\rangle\rangle_2 ,  \label{firstlinepsi2} \\
& = \frac{1}{\sqrt{2^N}} \sum_{m=0}^{L-1} \left(  \sum^N_{  \underset{  \{ k (\text{mod} L ) =m \} }{k=0} }   \sqrt{C_{N}^{k}}|k\rangle  \right)  |\frac{e^{2 \pi i m/L }}{\sqrt{2}},\frac{1}{\sqrt{2}}\rangle\rangle_2 .
\label{M2psiv2}
\end{align}
Here we made the replacement
\begin{align}
\sum_{k=0}^N \rightarrow \sum_{m=0}^{L-1} \sum^N_{  \underset{  \{ k (\text{mod} L ) =m \} }{k=0} } 
\end{align}
where the second summation only runs over those $ k $ satisfying $ k (\text{mod} L) = m $. 

It is then natural to define the set of states
\begin{align}
|{\cal C}_m \rangle = \frac{1}{\sqrt{Z_m}}  \sum^N_{  \underset{  \{ k (\text{mod} L ) =m \} }{k=0} }  \sqrt{C_{N}^{k}}|k\rangle 
\end{align}
where the normalization $ Z_m \approx 2^N/L $ is a good approximation for $ N \gg 1 $.  These states form an orthonormal set $ \langle {\cal C}_m | {\cal C}_n \rangle = \delta_{mn} $. Taking a Fourier transform of these states gives a spin coherent state
\begin{align}
\frac{1}{\sqrt{L}} \sum_{m=0}^{L-1} e^{-2 \pi i n m/L } |{\cal C}_m \rangle  =| \frac{e^{-2 \pi i n /L }}{\sqrt{2}}, \frac{1}{\sqrt{2}} \rangle \rangle . 
\end{align}
The inverse Fourier transform is
\begin{align}
|{\cal C}_m \rangle = \frac{1}{\sqrt{L}} \sum_{n=0}^{L-1} 
e^{2 \pi i n m/L } | \frac{e^{-2 \pi i n /L }}{\sqrt{2}}, \frac{1}{\sqrt{2}} \rangle \rangle .
\label{invfouriercat}
\end{align}
We see that $ |{\cal C}_m \rangle$ can be written as a sum of spin coherent states.  This is a spin-cat state which has been considered in various contexts before to encode quantum information \cite{agarwal1997atomic,semenenko2016implementing,qin2021generating,omanakuttan2024fault}.  

The $ M = 2$ state at the magic times can be written
\begin{align}
|\psi_2 (t= \frac{\pi}{2L} )  \rangle = \frac{1}{\sqrt{L}} \sum_{m=0}^{L-1} |{\cal C}_m \rangle_1 |\frac{e^{2 \pi i m/L }}{\sqrt{2}},\frac{1}{\sqrt{2}}\rangle\rangle_2 . 
\label{bellcat}
\end{align}
For well-separated spin coherent states
\begin{align}
|\langle \langle \frac{e^{2 \pi i n/L }}{\sqrt{2}},\frac{1}{\sqrt{2}} |  \frac{e^{2 \pi i m/L }}{\sqrt{2}},\frac{1}{\sqrt{2}}\rangle\rangle | & = \cos^N (n-m)\pi/L \nonumber \\
&  \approx e^{-N (n-m)^2 \pi^2/2 L^2} \nonumber \\
& \approx \delta_{nm}    \hspace{5mm} \text{(if $ L< \sqrt{N} $) .}
\label{orthogonalscs}
\end{align}
Hence we can consider the state (\ref{bellcat}) to be an entangled state of $ L $ terms where on the first ensemble there is a spin-cat state and on the second ensemble there is a spin coherent state at various positions on the equator of the Bloch sphere.  This fact was already noted in Ref. \cite{byrnes2013fractality} for the case $ L = 2$.

\section{Entanglement distribution at magic times}

We now return to the general case of $ M $ ensembles and evaluate the quantum state at the magic times.  Using the same methods as in the previous section we obtain the wavefunction before measurement for even $ M $ as
\begin{multline}
|\psi_M \rangle=\frac{1}{\sqrt{L^{\frac{M}{2}}}}\sum_{m_1,m_3,\ldots,m_{M-1}=0}^{L-1}  \\
\times | {\cal C}_{m_1}  \rangle_1  |\frac{e^{2 i(m_{1}-m_{3})\pi/L }}{\sqrt{2}},\frac{1}{\sqrt{2}}\rangle\rangle_2 | {\cal C}_{m_3} \rangle_3
|\frac{e^{2 i(m_{3}-m_{5})\pi/L }}{\sqrt{2}},\frac{1}{\sqrt{2}}\rangle\rangle_4  \\
\dots \otimes  | {\cal C}_{m_{M-1}} \rangle_{M-1} |\frac{e^{2 i m_{M-1}\pi/L }}{\sqrt{2}},\frac{1}{\sqrt{2}}\rangle\rangle_{M} . 
\label{evenpsiMcat}
\end{multline}
Written in this form we may deduce why there is the entanglement invariance property that was observed in Fig. \ref{fig:entanglement}.  Let us make the measurement of the intermediate ensembles in two steps, first projecting the even numbered nodes within $ j \in [2,M-1] $, then the odd nodes. Splitting the measurements into two steps does not affect the final result as all the measurements commute.  Again set the rotation parameter $ \phi = 0 $ for now.  Considering first for 
simplicity obtaining the outcome 
\begin{align}
|q= N \rangle^{(x)} = | \frac{1}{\sqrt{2}}, \frac{1}{\sqrt{2}} \rangle \rangle 
\end{align}
and assume that $L < \sqrt{N} $ such that the spin coherent states are approximately orthogonal as in (\ref{orthogonalscs}).  Then this enforces $ m_1 = m_3 = \dots = m_{M-1} $ for even $ M $.  The resulting state for even $ M $ is approximately
\begin{align}
\bigotimes_{j=2,4,\dots M-2} |q= N \rangle_j^{(x)} \langle q = N |_j^{(x)}  |\psi_M \rangle \propto \nonumber \\
\sum_{m=0}^{L-1} | {\cal C}_{m}  \rangle_1 | {\cal C}_{m}  \rangle_3 \dots | {\cal C}_{m}  \rangle_{M-1} |\frac{e^{2 i m \pi/L }}{\sqrt{2}},\frac{1}{\sqrt{2}}\rangle\rangle_{M} ,  
\end{align}
which is valid for $ N \gg 1 $.  
This correlates all the remaining ensembles, in particular the first and last ensemble such that they have the same label $ m $. 

Now looking at the remaining intermediate ensembles, the spin-cat states can be written as a superposition of spin coherent states as shown in (\ref{invfouriercat}).  Measurement of the cat states will collapse the superposition as
\begin{align}
\langle q = N |^{(x)}  | {\cal C}_{m}  \rangle \approx \frac{1}{\sqrt{L}}
\end{align}
where only the $ n = 0 $ term in (\ref{invfouriercat}) contributes for well-separated spin coherent states.  The correlations between the first and last ensembles are hence preserved and we obtain
\begin{align}
 \bigotimes_{j=2}^{M-1} & |q= N \rangle_j^{(x)} \langle q = N |_j^{(x)}  | \psi_M \rangle \propto \nonumber \\
& \sum_{m=0}^{L-1} | {\cal C}_{m}  \rangle_1 |\frac{e^{2 i m \pi/L }}{\sqrt{2}},\frac{1}{\sqrt{2}}\rangle\rangle_{M}  , 
\label{projideal}
\end{align}
which is valid for $ N \gg 1 $.  
This is precisely the same quantum state as (\ref{bellcat}).  This is obtained for arbitrary chain length $ M $.   This explains the invariance property of the entanglement and fidelity observed in Figs. \ref{fig:entanglement} and \ref{fig:fidelity} for the even case. 

For odd $ M $, we can use the same manipulations to obtain
\begin{align}
 \bigotimes_{j=2}^{M-1} & |q= N \rangle_j^{(x)} \langle q = N |_j^{(x)}  | \psi_M \rangle \propto  \sum_{m=0}^{L-1} | {\cal C}_{m}  \rangle_1 | {\cal C}_{m}  \rangle_{M} . 
\label{projideal}
\end{align}
This is the same state as (\ref{ideal_odd}) which can be written as
\begin{align}
| \Psi_N^{(M=3)}  \rangle & \approx  \frac{1}{\sqrt{L}} \sum_{m=0}^{L-1} | {\cal C}_{m}  \rangle_1 | {\cal C}_{m}  \rangle_{3} , \\
& =  \frac{1}{\sqrt{L}} \sum_{n=0}^{L-1} |\frac{e^{2 i n \pi/L }}{\sqrt{2}},\frac{1}{\sqrt{2}}\rangle\rangle_{1} |\frac{e^{-2 i n \pi/L }}{\sqrt{2}},\frac{1}{\sqrt{2}}\rangle\rangle_{3} 
\end{align}
which is valid for $ N \gg 1 $.

\section{Measurements of spin coherent states}

In the previous section we showed that for the measurement outcome $ q_j = N $ at the magic times, an entangled state consisting of spin-cat states is generated.  
We now wish to generalize this result for arbitrary measurement outcomes. One can see from the form of (\ref{evenpsiMcat}) that if one were able to measure in the spin coherent state basis, a similar result would be obtained for all measurement outcomes.  This is however experimentally challenging, since unlike the optical case, homodyne measurements are not easily realizable with spin ensemble systems.  Therefore, an alternative is required to implement the measurements of the last section.  In this section we show that standard spin measurements are sufficient in achieving the same collapse as described in the previous section.  

\begin{figure}
    \centering
    \includegraphics[width=\columnwidth]{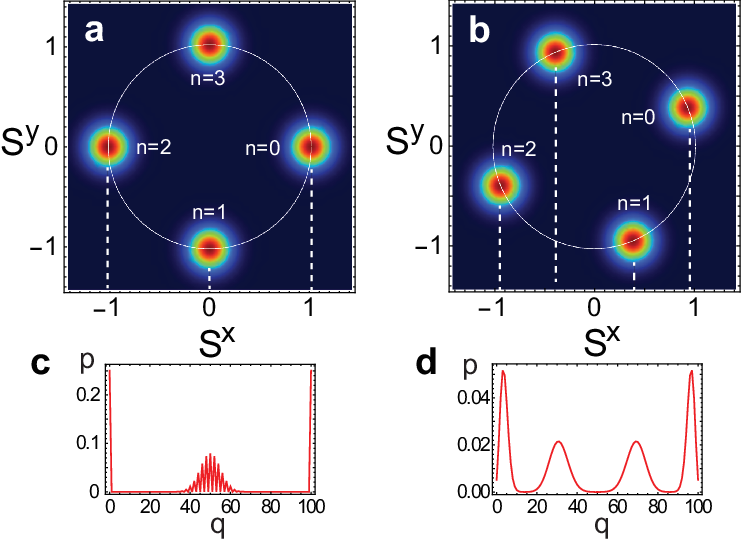}
    \caption{Visualization and measurement of a spin-cat state. (a)(b) Visualization of the spin-cat states (\ref{invfouriercat}) and (\ref{rotatedspincat}) respectively. Figures are plotted by centering a Gaussian $ e^{- N[ (\langle S^x \rangle - \cos \theta)^2 + (\langle S^y \rangle - \sin \theta)^2]/4} $, which is the Q-function distribution for a spin coherent state.  (c)(d) The probability distribution (\ref{probcat}) for $ \phi = 0, \pi/2L $ respectively.  For all plots $ L = 4, N = 100 $ is used.  }
    \label{fig:measurement}
\end{figure}

To illustrate the measurement scheme, first consider performing a measurement of the spin-cat state (\ref{invfouriercat}) in the $ x $ basis.  Plotting the positions of the spin coherent states involved in this superposition immediately identifies a problem: spin coherent states labeled with $ n $ and $ L-n $ have the same $ x $ position (Fig. \ref{fig:measurement}(a)).  Hence projection in the $ x $ basis will collapse the superposition, but will catch two spin coherent states together, unless they are located at the extremal positions $ \langle S^x \rangle = \pm N $.  To avoid this issue, it is beneficial to slightly offset the spin coherent states such that one measures instead the state
\begin{align}
e^{i \phi n^a } |{\cal C}_m \rangle = \frac{1}{\sqrt{L}} \sum_{n=0}^{L-1} 
e^{2 \pi i n m/L } | \frac{e^{-i (2 \pi n /L - \phi) }}{\sqrt{2}}, \frac{1}{\sqrt{2}} \rangle \rangle ,
\label{rotatedspincat}
\end{align}
where a suitable angle choice is
\begin{align}
\phi = \frac{\pi}{2L }.  
\end{align}
This slight rotation offsets the spin coherent states slightly such that a projection on the $ x $-axis results in a collapse of the state to uniquely one of the spin coherent states (Fig. \ref{fig:measurement}(b)). 

The effect of the measurement is also evident by examining the probability of the projection
\begin{align}
p_q =| \langle q |^{(x)}  e^{i \phi n^a } |{\cal C}_m \rangle|^2 .
\label{probcat}
\end{align}
Figure \ref{fig:measurement}(c)(d) shows the probability of the rotated spin-cat state for the same states as shown in Figs. \ref{fig:measurement}(a)(b) respectively. 
We see peaks in the probability distribution at the expectation value of $ S^x $ each of the spin coherent states involved in the spin-cat state. When the projection involves two spin coherent states, there is a rapid oscillation of the probability distribution (Fig. \ref{fig:measurement}(c)).  By applying the rotation $ \phi $ this ambiguity is removed and smooth Gaussian peaks are seen for each of the spin coherent states (Fig. \ref{fig:measurement}(d)). 

For a given measurement outcome $ q $, one can deduce which spin coherent state has been projected using the peak position of the Gaussians:
\begin{align}
q_{\text{peak}} = N \cos^2 ( \pi ( \bar{n}-1/4)/L )  , 
\end{align}
where the spin coherent state is labeled by $ \bar{n} \in [0, L-1] $. For a particular outcome $ q $, by finding the closest $ q_{\text{peak}} $ one can deduce which spin coherent state labeled by $ \bar{n} $ the collapse occurred at.

\section{Approximate wavefunction for arbitrary measurement outcomes}
\label{sec:approxwav}

Using the technique in the previous section, we may now examine the remaining projection outcomes in addition to the $ q_j = N $ case.  Examining  (\ref{evenpsiMcat}), first consider performing a measurement on the even numbered ensembles.  Assuming that the collapse occurs such as to pick out one and only one of the spin coherent states we obtain the relation
%
\begin{align}
    m_{j-1} - m_{j+1} = \bar{m}_j (\text{mod} L) .  
\end{align}
The state is then
\begin{align}
& \bigotimes_{j=2,4,\dots M-2} |q_j \rangle_j^{(x)} \langle q_j |_j^{(x)}  |\psi_M \rangle \propto \nonumber \\
& \sum_{m=0}^{L-1} | {\cal C}_{m}  \rangle_1 | {\cal C}_{m - \bar{m}_2}  \rangle_3 | {\cal C}_{m - \bar{m}_2 - \bar{m}_4}  \rangle_5 \dots | {\cal C}_{m - \bar{m}_{\text{even}}}  \rangle_{M-1} \nonumber \\
 & \otimes |\frac{e^{2 i (m+ \bar{m}_{\text{even}})  \pi/L }}{\sqrt{2}},\frac{1}{\sqrt{2}}\rangle\rangle_{M}  .  
 \label{evenprojectedpsi}
\end{align}
where we defined
\begin{align}
\bar{m}_\text{even} =  
\sum_{j=2,4, \dots, M-2} \bar{m}_j .
\end{align}
This correlates the first and last ensemble up to an offset in $ m $. In (\ref{evenprojectedpsi}), the proportionality factor is
\begin{align}
& \frac{1}{\sqrt{L^{M/2}}} \prod_{j=2,4,\dots, M-2} \langle q_j|^{(x)} | \frac{e^{2i (m_{j-1} - m_{j+1} ) \pi/L + i \phi }}{\sqrt{2}}, \frac{1}{\sqrt{2}} \rangle \rangle \nonumber \\
& = \frac{e^{iN \bar{m}_\text{even} \pi /L + iN \phi/2} }{\sqrt{L^{M/2}}} \nonumber \\
& \times \prod_{j=2}^{M-2}  \sqrt{C_N^{q_j}} \cos^{q_j} ( \frac{\bar{m}_j \pi}{L} + \frac{\phi}{2} )\sin^{N-q_j} ( \frac{\bar{m}_j \pi}{L} + \frac{\phi}{2} ) 
\end{align}
which has no dependence on $ m $ hence does not affect the wavefunction.  

Measurement of the odd numbered intermediate ensembles will give factors
\begin{align}
\langle q|^{(x)} | {\cal C}_m \rangle & = \frac{1}{\sqrt{L}} \sum_{n=0}^{L-1} e^{2 \pi i nm/L } 
\langle q|^{(x)}  | \frac{ e^{-2 \pi i n/L + i \phi}}{\sqrt{2}}, \frac{ 1}{\sqrt{2}} \rangle \rangle \nonumber \\
& \approx \frac{1}{\sqrt{L}} e^{i \pi \bar{n} (2m - N)/L + i \phi N/2 } \sqrt{C_N^{q}} \nonumber \\
& \times \cos^{q} ( \frac{\bar{n} \pi}{L} - \frac{\phi}{2} )\sin^{N-q_j} ( \frac{\bar{n} \pi}{L} - \frac{\phi}{2} ) 
\end{align}
where in the second line we used the result of the previous section that the measurement will collapse the spin-cat state to one of the spin coherent states with label $ \bar{n} $.  Here the important factor that affects the wavefunction is the phase dependence on $ m $.  The total $ m $ dependence is given by projecting the odd sites is
\begin{align}
\langle q_3|^{(x)} | {\cal C}_{m -\bar{m}_2} \rangle 
\langle q_5|^{(x)} | {\cal C}_{m -\bar{m}_2 - \bar{m}_4 } \rangle \dots 
\langle q_{M-1} |^{(x)} | {\cal C}_{m - \bar{m}_\text{even} } \rangle \nonumber \\
\propto e^{2 \pi i m \bar{n}_{\text{odd}}/L }
\end{align}
where 
\begin{align}
\bar{n}_{\text{odd}} = \sum_{j=3,5,\dots, M-1} \bar{n}_j  . 
\end{align}

We thus finally obtain the final state after projection 
\begin{align}
& | \Psi_{\vec{q}}^{\text{approx}} \rangle =  \bigotimes_{j=2}^{M-1}  |q_j \rangle_j^{(x)} \langle q_j |_j^{(x)}  |\psi_M \rangle \propto \nonumber \\
& \sum_{m=0}^{L-1} e^{2 \pi i m \bar{n}_{\text{odd}}/L } | {\cal C}_{m}  \rangle_1 |\frac{e^{2 i (m - \bar{m}_{\text{even}})  \pi/L }}{\sqrt{2}},\frac{1}{\sqrt{2}}\rangle\rangle_{M} .  
 \label{evenprojectedpsiapprox}
\end{align}

\begin{figure}
    \centering
    \includegraphics[width=\columnwidth]{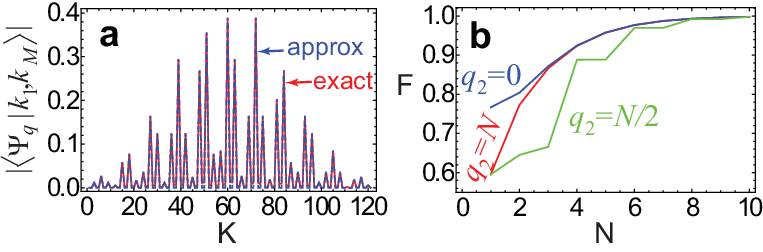}
    \caption{Comparison of the exact wavefunction (\ref{final_even}) with the approximate wavefunction (\ref{evenprojectedpsiapprox}).  (a) The wavefunction amplitude $ | \langle \Psi_{\vec{q}} | k_1, k_M \rangle | $ for $N = 10, M= 3, L = 3 $ and $ q_2 = N/2 $ for $ K = (N+1) k_1 + k_M +1 $.  The approximate wavefunction is shown as the dashed line, the exact wavefunction is shown as a solid line. (b) The fidelity $ F = |\langle \Psi_{\vec{q}} | \Psi_{\vec{q}}^{\text{approx}} \rangle |^2 $ for $ M = 3, L = 3$ for three outcomes $ q_2 = 0, N/2, N $ as marked. All cases implement the measurement scheme where the intermediate ensembles are measured using the offset angle $ \phi = \pi/2L $.
    }
    \label{fig:finalfig}
\end{figure}

Figure \ref{fig:finalfig} show a comparison of the approximate wavefunction (\ref{evenprojectedpsiapprox}) with the exact wavefunction (\ref{final_even}).  In Fig. \ref{fig:finalfig}(a) the amplitude of the coefficients in the number basis is shown.  Excellent agreement is seen, with the coefficents having nearly perfect agreement.  The fidelity of the approximate wavefunction improves with $ N $, as seen in Fig. \ref{fig:finalfig}(b).  The reason that the fidelities improve with $ N $ is due to the measurement scheme that we implement as described in the previous section.  In order to collapse the state to a single spin coherent state, the Gaussians for each one should be well-separated, as discussed in Fig. \ref{fig:measurement}.  This requires $ L < \sqrt{N} $ typically. With increasing $ N $, the Gaussian become better separated and approaches the ideal collapse to a single spin coherent state.  

We note that in Fig. \ref{fig:fidelity} perfect fidelity $ F = 1$ was obtained, whereas in Fig. \ref{fig:finalfig} only near-unit fidelity is obtained.  The reason is that in Fig. \ref{fig:fidelity} the offset angle $ \phi = 0 $ and only the outcome $ q_j = N $ was considered.  For this case, the projection on the spin coherent state is perfect, as can be seen in Fig. \ref{fig:measurement}(a) and \ref{fig:measurement}(c).  However, there is no guarantee that the outcome $q_j = N$ is obtained, in fact for long chains it has an exponentially small probability.  Using the offset measurement of Fig. \ref{fig:measurement}(b)(d) is advantageous in this sense, as all measurement outcomes lead to the generation of a nearly ideal state (\ref{evenprojectedpsiapprox}).  The small price to pay for this is a non-unit fidelity, although as can be seen in Fig. \ref{fig:finalfig}(b) the fidelity quickly approaches 1.

Therefore, using the offset measurement include $ \phi = \pi/2L$, one may obtain consistently obtain (\ref{evenprojectedpsiapprox}) for all measurement outcomes.  The final step to turn this into a deterministic scheme for preparing entanglement is to peform conditional rotations to eliminate the $ \bar{n}_\text{odd} $ and  $ \bar{m}_\text{even} $ terms.  Specifically, knowledge of $ \bar{n}_\text{odd} $ and  $ \bar{m}_\text{even} $  allows for a correction of the state to the state (\ref{projideal}) using the rotations
\begin{align}
e^{i 2 \pi n^a /L } | {\cal C}_m \rangle & = e^{i 2 \pi m /L } | {\cal C}_m \rangle \label{correction1} \\
e^{i \phi n^a } |\frac{e^{ i \phi' }}{\sqrt{2}},\frac{1}{\sqrt{2}}\rangle\rangle & = |\frac{e^{ i (\phi+\phi')  }}{\sqrt{2}},\frac{1}{\sqrt{2}}\rangle\rangle .
\label{correction2} 
\end{align}
Then by applying the corrections
\begin{align}
e^{-2 \pi i \bar{n}_{\text{odd}}  n^a_1/L }  e^{2 \pi i \bar{m}_{\text{even}}  n^a_M/L }  | \Psi_{\vec{q}} \rangle \approx | \psi_2 (t= \frac{\pi}{2L} ) \rangle 
\label{correction3}
\end{align}
we obtain a deterministic preparation of the state (\ref{bellcat}) which is the aim of this study.

\section{Resource Scaling and Experimental Implementation}

We now comment on the resource costs required to execute the protocol.  This question is intimately connected to the experimental implementation, since the resources required depend on the way that the operations are implemented. 
The key aspect that makes this scheme efficient is that the protocol has no dependence on $ N $, the number of qubits in the ensembles, in any of the operations that need to be performed.  Despite this, the amount of entanglement that is distributed is of order $ \log_2 N $, as can be seen in Fig. \ref{fig:entanglement}.  The reason for this is that only collective spin operations are used to control the ensembles in all the steps of the protocol. 

To make this point clear, let us review the steps of the protocol and examine them each:
\begin{enumerate}
\item Prepare the initial state $ | \phi_0 \rangle = |\frac{1}{\sqrt{2}}, \frac{1}{\sqrt{2}} \rangle \rangle^{\otimes M} $. 
\item Apply the Hamiltonian (\ref{mainham}) for a time $ t $. 
\item Apply the rotation (\ref{equatorrot}) to the intermediate qubits. 
\item Measure the intermediate qubits in the $ x $-basis (\ref{measurementop}). 
\item Based on the measurement results, apply the correction (\ref{correction3}).  
\end{enumerate}
Specifically, the initial state preparation in Step 1 can be routinely performed in atomic ensembles using optical pumping to populate one of the atomic levels, followed by a $ \pi/2 $ pulse implemented with radio frequency (RF) pulses. Both of these operations are independent of $ N $ since the optical and RF pulses applied to the ensembles are common to all atoms in the ensemble.  The unitary operations in Steps 2, 3, 5 are all in terms of total spin operators $ S^\gamma $, hence are at the ensemble level. Again these can be implemented using RF pulses on the hyperfine ground states of the atoms.  The $ S^z S^z $ interaction of Step 2 can be implemented using techniques discussed in Ref. \cite{pyrkov2013entanglement,hussain2014geometric}, where the interaction is mediated using an optical channel.  Finally, the measurement in Step 4 is performed using atomic counting of the ensemble, which is routinely performed in works such as Refs. \cite{Hume2013,Ott2016,Qu2020}.  All these operations only require operations at the ensemble level, not at the microscopic level of the individual qubits.  For this reason there is no $ N $ dependence to the scheme.  The only scaling is with respect to the chain length $ M $, where the $ M - 1 $ links of the chain must be entangled and $ M - 2$ intermediate nodes must be projected out.

\section{Conclusions}

We have proposed a protocol to distribute macroscopic entanglement on ensembles of qubits, by first generating entanglement in a chain configuration then projecting out the intermediate ensembles to leave the ends of the chain entangled.  Such a protocol is non-trivial in the ensemble case since the entanglement is of a complex many-body nature, which causes the entanglement tends to diminish with longer chain lengths.  We have found ``magic'' interaction times where the entanglement can be transferred without degradation.  The nature of the states at these magic times was found due to an encoding of the entangled states in terms of generalized spin-cat states.  This allowed us to derive the explicit form of the states for any ensemble size. The final wavefunction (\ref{final_even}), which has the approximate form (\ref{evenprojectedpsiapprox}), takes a similar form to the $ M = 2 $ ensemble case, up to simple spin rotations.  By correcting for these known rotations, it allows for the deterministic preparation of the equivalent state generated by an $ S^z S^z $ interaction.  

The amount of entanglement that is generated using this scheme is of order $ \log_2 N$, and hence is macroscopic by our definition.   The resources required to distribute this do not involve $ N $, and only depend upon $ M $, the chain length.  The reason is that only total spin operations are used in the protocol from start to finish. In this sense this has a considerable scaling advantage over preparing the entanglement microscopically.  The parallelization occurs since the resources to control an ensemble are the same as that of a qubit. The use of total spin operators means that the control pulses illuminate the entire ensemble simultaneously, which is the origin of the parallelization.  The type of entanglement that is distributed is suitable for various protocols that are based on the $ S^z S^z $ interaction, since this is ultimately the form of entanglement that is distributed.  This includes schemes such as quantum teleportation \cite{pyrkov2014quantum,pyrkov2014full}, Deutsch-Jozsa algorithm \cite{semenenko2016implementing}. Such tasks could be implemented on a distributed quantum computing platform.  The distribution of entanglement at macroscopic quantities is a task that we anticipate will become more important as the quantum internet gains more prominence.  Our protocol offers the possibility of distribute entanglement at scale, which can serve as a fundamental primitive in quantum network infrastructure.



\section{Acknowledgments}

A. N. P and I. D. L.  are supported by the Russian Science Foundation (grant no 23-21-00507). T.B. is supported by the National Natural Science Foundation of China (62071301); NYU-ECNU Institute of Physics at NYU Shanghai; Shanghai Frontiers Science Center of Artificial Intelligence and Deep Learning; the Joint Physics Research Institute Challenge Grant; the Science and Technology Commission of Shanghai Municipality (19XD1423000,22ZR1444600); the NYU Shanghai Boost Fund; the China Foreign Experts Program (G2021013002L); the NYU Shanghai Major-Grants Seed Fund; Tamkeen under the NYU Abu Dhabi Research Institute grant CG008; and the SMEC Scientific Research Innovation Project (2023ZKZD55).

\appendix

\section{Expressions for odd $ M $}
\label{sec:oddMexpressions}

Here we show the corresponding expressions for odd $ M $ for the even $ M $ expressions given in the main text. 

The odd $ M $ counterpart of (\ref{final_even}) is 
\begin{multline}
\label{final_odd}
| \Psi_{\vec{q}} \rangle=\frac{1}{\sqrt{2^{\frac{(M+1)N}{2}}}}\sum_{k_1,k_3,\ldots,k_M=0}^{N} \sqrt{ C^{k_1}_{N}  C^{k_M}_{N} } 
 \\ \times \left( \prod_{j=2}^{M-1} \Omega_{q_j}^{(j)} \right) |k_1\rangle|k_M\rangle . 
\end{multline}
In order to calculate the fidelity for odd case, we take the overlap with the projected $ M = 3 $ state (\ref{ideal_odd}). The fidelity in this case equals 
\begin{multline}
\label{fidelity_odd}
F = \frac{| \langle  \psi_3 | \Psi_{\vec{q}} \rangle |^2}{p_{\vec{q}} }   =\frac{1}{2^{N(M+5)/2} p_{\vec{q}} }\\
\times \frac{\Big|\sum_{k_1, k_3, \dots, k_M } 
 C_N^{k_1}C_N^{k_M} {\Omega_{N}^{(2)}}^* \left( \prod_{j=2}^{M-1} \Omega_{q_j}^{(j)} \right)   \Big|^2 }{ \sum_{k_1,k_3=0}^{N}  C^{k_1}_{N}  C^{k_3}_{N} | \Omega_{N}^{(2)} |^2 }
\end{multline}
when we measure $N$ on all  intermediate ensembles.

The odd $ M $ counterpart of (\ref{evenpsiMcat}) is
\begin{multline}
| \psi_M \rangle=\frac{1}{\sqrt{L^{\frac{(M+1)}{2}}}}\sum_{m_1,m_3,\ldots,m_M=0}^{L-1} \\
\times | {\cal C}_{m_1}  \rangle_1  |\frac{e^{2 i(m_{1}-m_{3})\pi/L }}{\sqrt{2}},\frac{1}{\sqrt{2}}\rangle\rangle_2 | {\cal C}_{m_3} \rangle_3
|\frac{e^{2 i(m_{3}-m_{5})\pi/L }}{\sqrt{2}},\frac{1}{\sqrt{2}}\rangle\rangle_4  \\
\dots \otimes |\frac{e^{2 i(m_{M-2}-m_{M})\pi/L }}{\sqrt{2}},\frac{1}{\sqrt{2}}\rangle\rangle_{M-1} | {\cal C}_{m_{M}}  \rangle_{M} .
\label{oddpsiMcat}
\end{multline}
Following similar steps to that in Sec. \ref{sec:approxwav} we get the odd $ M $ counterpart of (\ref{evenprojectedpsiapprox})
\begin{align}
& | \Psi_{\vec{q}}^{\text{approx}} \rangle =  \bigotimes_{j=2}^{M-1}  |q_j \rangle_j^{(x)} \langle q_j |_j^{(x)}  |\psi_M \rangle \propto \nonumber \\
& \sum_{m=0}^{L-1} e^{2 \pi i m \bar{n}_{\text{odd}}/L } | {\cal C}_{m}  \rangle_1 | {\cal C}_{m - \bar{m}_{\text{even}}}  \rangle_M  .  \label{oddprojectedpsiapprox}
\end{align}
where for (\ref{oddprojectedpsiapprox}) we use the modified definitions
\begin{align}
\bar{m}_\text{even} & =  
\sum_{j=2,4, \dots, M-1} \bar{m}_j  \nonumber \\
\bar{n}_{\text{odd}} & = \sum_{j=3,5,\dots, M-2} \bar{n}_j  . 
\end{align}



\end{document}